\documentclass[11pt]{article}
\usepackage{amsmath}
\usepackage{amsfonts}
\usepackage{amssymb}
\usepackage{graphicx}
\usepackage{bm}
\usepackage{color}
\usepackage{amscd}

\def\bea{\begin{eqnarray}}
\def\eea{\end{eqnarray}}

\begin{document}
\begin{center}
\LARGE {\bf The entropy formula of black holes in  Minimal Massive Gravity and its application for BTZ black holes}
\end{center}
\begin{center}
{\bf M. R. Setare\footnote{rezakord@ipm.ir} }\hspace{1mm} ,
H. Adami \footnote{E-mail: hamed.adami@yahoo.com}\hspace{1.5mm} \\
 {Department of Science, Campus of Bijar, University of  Kurdistan  \\
Bijar, IRAN.}
 \\
 \end{center}
\vskip 3cm

\begin{abstract}
In this paper we obtain the entropy formula of black hole solutions of  Minimal Massive Gravity (MMG) by Tachikawa method \cite{a}.
Then we apply this formula for BTZ black hole solution. We find that the usual Bekenstein-Hawking entropy is modified. The modification come from
Chern-Simons (CS) term and new term in MMG. The contribution of CS term, which is proportional with inner radius of horizon is not new, but last term which is due to the new term of MMG, as usual is proportional to the outer radius of horizon and is new result. Then we show that the total entropy
exactly can be reproduced by Cardy formula for entropy of the dual boundary CFT.
\end{abstract}

\newpage

\section{Introduction}
Recently an interesting three dimensional massive gravity introduced by Bergshoeff, et.al \cite{17} which dubbed Minimal Massive Gravity (MMG), which has the same minimal local structure as Topologically Massive Gravity (TMG) \cite{4}. The MMG model has the same gravitational
degree of freedom as the TMG has and the linearization of the metric field equations for MMG yield a single propagating
massive spin-2 field. So both models have the same spectrum \cite{18}. However, in contrast to TMG, there is not bulk vs boundary clash in the framework of this new model. One can obtain MMG  by adding the CS deformation term and an extra term
to pure Einstein gravity with a negative cosmological constant. So importance of the work \cite{17} is not only solve the problem of TMG, but also do this by introduction only one parameter. During last months some interesting works have been done on MMG model \cite{18}. More recently, this model has been extended to General Minimal Massive Gravity theory (GMMG) \cite{f}. GMMG is a unification of MMG with New Massive Gravity (NMG) \cite{6}, so this model is realized by adding the higher derivative deformation term to the Lagrangian of MMG.\\
We know that the BTZ \cite{e} black hole solution of Einstein gravity with a negative cosmological constant, which has constant curvature solves trivially TMG. In the other hand since MMG locally has the same structure as that of TMG, BTZ should also be solution of MMG. Now it is interesting
to obtain entropy of BTZ black hole as a solution of MMG and investigate the contribution of new term of MMG in entropy.\\
The general formula for entropy of black hole solution of arbitrary general covariant Lagrangian constructed from metric has been obtained by Wald, et.al \cite{g,g,i}. But one can not apply the Wald's formula to obtain entropy of black holes in the context of MMG, for couple reasons. At first, because the Lagrangian of MMG is not covariant due to the presence of CS term. Second, because the the Lagrangian of MMG has written in first formalism  due to the new term of MMG. The extension of the Wald's approach to the TMG, was provided by Tachikawa \cite{a} (see also \cite{d,j}).
In the present paper we show that Tachikawa method work for MMG as well. To apply this approach to the black hole solution of MMG, it is sufficient to obtain the conserved current and corresponding conserved charge. This formulation for conserved charge reproduce the expression of total entropy of black holes in MMG. Here we probe this method for BTZ solutions. The total entropy of BTZ solution has dual meaning in terms of boundary CFT. Then we show that the total entropy can be reproduce exactly by using the Cardy formula.

\section{ Minimal massive gravity and conserved charge}
The Lagrangian 3-form of MMG is given by \cite{17}
\begin{equation} \label{1}
 L_{MMG}=L_{TMG}+\frac{\alpha}{2}e.h\times h
\end{equation}
where $L_{TMG}$ is the Lagrangian of TMG,
 \begin{equation} \label{1'}
 L_{TMG}=-\sigma e.R+\frac{\Lambda_0}{6} e.e\times e+h.T(\omega)+\frac{1}{2\mu}(\omega.d\omega+\frac{1}{3}\omega.\omega\times\omega)
\end{equation}
 where $\Lambda_0$ is a cosmological parameter with dimension of mass squared, and $\sigma$ a sign. $\mu$ is mass parameter of Lorentz Chern-Simons term. $\alpha$ is a dimensionless parameter, $e$ is dreibein, $h$ is the auxiliary field, $\omega$ is dualised  spin-connection, $T(\omega)$ and
  $R(\omega)$ are Lorentz covariant torsion and curvature 2-form respectively.
One can rewrite the Lagrangian 3-form $L_{MMG}$ as following

$$ L_{MMG}=-\sigma e_a R^a+\frac{\Lambda_0}{6}\varepsilon^{abc}e_a e_b e_c+h_a T^a+\frac{1}{2\mu}[\omega_a d\omega^a+\frac{1}{3}\varepsilon^{abc}
 \omega_a \omega_b \omega_c]$$
 \begin{equation} \label{22'}
 +\frac{\alpha}{2}\varepsilon^{abc}e_a h_b h_c
\end{equation}
The equations of motion of the above Lagrangian by making variation with respect to the fields $h$, $e$ and $\omega$  are as following respectively

\begin{equation} \label{EOM1}
T(\omega)+\alpha e\times h=0
\end{equation}
\begin{equation} \label{EOM2}
-\sigma R(\omega)+\frac{\Lambda_0}{2}e\times e+D(\omega)h+\frac{\alpha}{2}h\times h=0
\end{equation}
\begin{equation} \label{EOM3}
R(\omega)+\mu e\times h-\sigma\mu T(\omega)=0
\end{equation}
where the locally Lorentz covariant torsion and curvature 2-forms are
\begin{equation} \label{EOM5}
T(\omega)=de+\omega\times e,
\end{equation}
\begin{equation} \label{EOM4}
R(\omega)=d\omega+\frac{1}{2}\omega\times\omega
\end{equation}
The covariant exterior derivative $D(\omega)$ in Eq.(\ref{EOM2}) is given by
\begin{equation} \label{EOM6}
D(\omega)h=dh+\omega\times h
\end{equation}
The first-order variation of $L_{MMG}$ is given by

$$ \delta L_{MMG} = \delta e \cdot [  -\sigma R(\omega) +\frac{1}{2} \Lambda _{0} e \times e +D(\omega) h+\frac{\alpha}{2} h\times h ] $$
$$ + \delta \omega \cdot [ \frac{1}{\mu} R(\omega) - \sigma T(\omega) + e \times h]
+\delta h \cdot [ T(\omega) + \alpha e \times h] $$
\begin{equation}\label{2}
 +d ( -\sigma \delta \omega \cdot e +\frac{1}{2 \mu} \delta \omega \cdot \omega +\delta e \cdot h ) ,
\end{equation}
where we have used the relations $[\delta, d]\omega=0$, $[\delta, d]e=0$. We can express the above variation of $L_{MMG}$ as \cite{a}
\begin{equation}\label{3'}
\delta L_{MMG}=\sum_{\Phi}E_{\Phi}\delta\Phi+d\Theta(\Phi, \delta\Phi)
\end{equation}
here $E_{\Phi}$ denotes the equation of motion associated with field $\Phi$, but $\Theta$ is sympletic potential. From Eq.(\ref{2})  sympletic potential is given by
\begin{equation}\label{3}
\Theta = -\sigma \delta \omega \cdot e +\frac{1}{2 \mu} \delta \omega \cdot \omega +\delta e \cdot h .
\end{equation}
Now we consider following condition as has been considered by Tachikawa in \cite{a} for TMG,
 \begin{equation}\label{4'}
\delta_{\xi} L_{MMG}(\Phi)=\pounds_{\xi}L_{MMG}(\Phi)+d\Psi_{\xi}
\end{equation}
where $\Psi_{\xi}$ is a suitable 2-form, $\delta_{\xi}$ is the variation induced by diffeomorphism by the vector field $\xi$, while $\pounds_{\xi}$
denotes the Lie derivative with respect to $\xi$. As have been mentioned in \cite{a} the above condition means that the diffeomorphism generated
 by $\xi$ is a symmetry, the corresponding conserved on-shell is \cite{b}
 \begin{equation}\label{44'}
J_{\xi} = \Theta (\Phi , \delta _{\xi} \Phi) -i_{\xi}L_{MMG}-\Psi_{\xi} ,
\end{equation}
so, $dJ_{\xi}\simeq 0$, where $\simeq$ denotes the equality on-shell. $i_{\xi}$ is the interior product. $J_{\xi}$ is exact on-shell, so there
is a 1-form $Q_{\xi}$, such that \cite{c}
\begin{equation}\label{4}
J_{\xi} \simeq d Q_{\xi}
\end{equation}
Thus $J_{\xi}$ is the conserved  Noether current and $Q_{\xi}$ is corresponding conserved charge.\\
In our problem, since $\omega$ and $e$ are tensor so
\begin{equation}\label{5'}
\delta_{\xi}\omega=\pounds_{\xi}\omega \hspace{0.5cm} \delta_{\xi}e=\pounds_{\xi}e
\end{equation}
due to this, in Eq.(\ref{44'}) $\Psi_{\xi}=0$ and we have
\begin{equation}\label{444'}
J_{\xi} = \Theta (\Phi , \delta _{\xi} \Phi) -i_{\xi}L_{MMG}.
\end{equation}
By substituting $\Theta$ and $L_{MMG}$ from Eqs.(\ref{3}), (\ref{1}) respectively into Eq.(\ref{444'}), and replacing variation $\delta_{\xi}\omega$
, $\delta_{\xi}e$ by their Lie derivative and also using following identity
\begin{equation}\label{55'}
\pounds_{\xi}=di_{\xi}+i_{\xi}d
\end{equation}
finally we obtain following expression  for $Q_{\xi}$
\begin{equation}\label{5}
Q_{\xi} = -\sigma i_{\xi} \omega \cdot e +\frac{1}{2 \mu} i_{\xi} \omega \cdot \omega +i_{\xi} e \cdot h .
\end{equation}
Now by inserting $T(\omega)$ from Eq.(\ref{EOM5}) into equation of motion (\ref{EOM1}) we obtain \cite{17}
\begin{equation}\label{8}
de+\Omega \times e =0,
\end{equation}
where new dual spin-connection 1-form $\Omega (e)$ which is torsion-free is given by
\begin{equation}\label{8'}
 \Omega =\omega +\alpha h
\end{equation}
 By considering $ 1+\alpha \sigma \neq 0 $, and using following identities
 \begin{equation}\label{88'}
 D(\Omega)T(\Omega)\equiv R(\Omega)\times e, \hspace{0.5cm}D(\Omega)R(\Omega)\equiv0,
\end{equation}
  one finds that the field equations imply \cite{17}
  \begin{equation}\label{888'}
  e \cdot h = 0
\end{equation}
  then as have been obtained in \cite{17} we have
\begin{equation}\label{9}
h_{\mu \nu} = -\frac{1}{\mu (1+\alpha \sigma)^{2}} \left( S_{\mu \nu}+ \frac{1}{2} \alpha \Lambda _{0} g_{\mu \nu} \right) .
\end{equation}
where  3D Schouten tensor $ S_{\mu \nu} $ is given by
\begin{equation}\label{10}
S_{\mu \nu} = R_{\mu \nu} - \frac{1}{4} g_{\mu \nu} R,
\end{equation}
Now by using Eq.(\ref{8'}) we rewrite $Q_{\xi}$ in Eq.(\ref{5}) as following
$$ Q_{\xi} = -\sigma i_{\xi} \Omega \cdot e +\frac{1}{2 \mu} i_{\xi} \Omega \cdot \Omega + i_{\xi} e \cdot h+\alpha\sigma i_{\xi}h\cdot e$$
\begin{equation}\label{11'}
-\frac{\alpha}{2\mu} \left(  i_{\xi} \Omega \cdot h + i_{\xi} h \cdot \Omega - \alpha  i_{\xi} h \cdot h \right) ,
\end{equation}
then using Eq.(\ref{888'}) the above charge takes final form
$$ Q_{\xi} = -\sigma i_{\xi} \Omega \cdot e +\frac{1}{2 \mu} i_{\xi} \Omega \cdot \Omega +(1+\alpha \sigma) i_{\xi} e \cdot h $$
\begin{equation}\label{11}
-\frac{\alpha}{2\mu} \left(  i_{\xi} \Omega \cdot h + i_{\xi} h \cdot \Omega - \alpha  i_{\xi} h \cdot h \right) ,
\end{equation}
The above equation give us the general conserved charge for calculations of entropy of any black hole solution of MMG.
\section{General formula for entropy and its application in BTZ black hole}
In this section we consider the general formula which has been obtained by Tachikawa \cite{a} (see also \cite{d}) for the Chern-Simons contribution to the entropy
\begin{equation}\label{12}
S=\frac{2\pi}{\kappa} \int _{\Sigma} Q_{\xi},
\end{equation}
where $\Sigma$ is the Cauchy surface, $\kappa$ is surface gravity, and $Q_{\xi}$
is conserved charge which we have obtained by equation (\ref{11}) in the context of MMG. The above entropy formula is a general equation for any black hole solution in the context of MMG, because as we have mentioned in the last paragraph of previous section, the conserved charge $Q_{\xi}$ is generic.\\
As an application of above formula here we consider the BTZ black hole solution, and probe this equation for such  black hole solutions of MMG.
Since MMG locally has the same structure as that of TMG then BTZ black hole solution of TMG, are also the solution of MMG. The metric of rotating
BTZ black hole is given by \cite{e}
\begin{equation}\label{12'}
ds^2=-\frac{(r^2-r_{+}^2)(r^2-r_{-}^2)}{l^2r^2}dt^2+\frac{l^2r^2}{(r^2-r_{+}^2)(r^2-r_{-}^2)}dr^2+r^2(d\phi-\frac{r_+r_-}{lr^2}dt)^2
\end{equation}
where $r_+$ and $r_-$ are outer and inner radiuses of horizon respectively. Now using following components of dreibein for the above metric
$$ e^{0}=\left( \frac{(r^{2}-r_{+}^{2})(r^{2}-r_{-}^{2})}{l^{2}r^{2}} \right)^{\frac{1}{2}}dt $$
$$ e^{1}= r \left( d \phi -\frac{r_{+}r_{-}}{lr^{2}} dt \right)  $$
\begin{equation}\label{13}
e^{2}=\left( \frac{l^{2}r^{2}}{(r^{2}-r_{+}^{2})(r^{2}-r_{-}^{2})} \right)^{\frac{1}{2}}dr ,
\end{equation}
and Kiling vector field
\begin{equation}\label{14}
\xi ^{\mu} = (1,0,\frac{r_{-}}{l r_{+}}) ,
\end{equation}
we obtain following expression for conserved charge $Q_{\phi}$ Eq.(\ref{11})
\begin{equation}\label{14'}
Q_{\phi}=\frac{(r_{+}^{2}-r_{-}^{2})}{l^2 r_+}(-\sigma r_+-\frac{r_-}{\mu l}-\frac{\alpha(1-\alpha \Lambda_0 l^2)r_+}{2\mu^2 l^2(1+\alpha\sigma)^2})
\end{equation}
Surface gravity $\kappa$ for BTZ metric is as
\begin{equation}\label{141}
 \kappa =\frac{r_{+}^{2}-r_{-}^{2}}{l^{2} r_{+}},
\end{equation}
Then by substituting Eqs.(\ref{14'}), (\ref{141}) into general entropy formula (\ref{12}), we obtain
\begin{equation}\label{15}
S = \frac{2\pi}{\kappa}\int_{0}^{2\pi} Q_{\phi} d\phi= -4 \pi ^{2} \sigma r_{+} - \frac{4 \pi ^{2} r_{-}}{\mu l}
- \frac{2\alpha \pi ^{2} r_{+} (1-\alpha \Lambda _{0} l^{2})}{\mu^{2} l^{2} (1+\alpha \sigma)^{2}},
\end{equation}
In order to obtain final form of entropy we should multiply the above result in factor $\frac{-1}{8\pi G}$, which is a coefficient in Lagrangian we have ignored at first (see also \cite{k}). So our final result for entropy of BTZ black hole solutions of MMG is as
\begin{equation}\label{152}
S=\frac{\pi}{2G}(\sigma r_++\frac{r_-}{\mu l}+\frac{\alpha r_+(1-\alpha\Lambda_0 l^2)}{2\mu^2 l^2(1+\alpha\sigma)^2})
\end{equation}
The first term in the above formula come from the Einstein-Hilbert and cosmological terms of Lagrangian $L_{MMG}$, the second term is the contribution of Chern-Simons term, and last term is due to the new term of MMG, which is a new result. It is interesting that by this method not only we have obtained the contribution of non-covariant part of Lagrangian (Chern-Simons term) in the entropy, but also we have obtained the contributions of another part as well. In the next section we show that the result (\ref{152}) for the entropy of BTZ solution of MMG is exactly coincide with what we obtain by Cardy formula for microscopic entropy of dual CFT of black hole.
\section{CFT consideration}
In this section we would like to compute the entropy of the BTZ black hole solution of MMG by assuming the existence of a dual CFT \cite{17}.The
left and right temperatures of the BTZ black hole is given by \cite{19} (see also \cite{20,21})
 \begin{equation}\label{35}
T_L=\frac{r_+-r_-}{2\pi l^2}, \hspace{0.5cm} T_R=\frac{r_++r_-}{2\pi l^2}
\end{equation}
In term of these quantities, the microscopic Cardy formula for the entropy is
 \begin{equation}\label{36}
S=\frac{\pi^2 l}{3}(C_L T_L+C_R T_R)
\end{equation}
 where $C_L$ and $C_R$ are left and right central charges of dual CFT respectively. As have been discussed in \cite{17}, the two copies of the
 Virasoro algebra on the boundary has the following central charges
 \begin{equation}\label{37}
C_L=\frac{3l}{2G}(\sigma-\frac{1}{\mu l}+\frac{\alpha(1-\alpha\Lambda_0 l^2)}{2\mu^2 l^2(1+\alpha\sigma)^2}), \hspace{0.3cm}C_R=\frac{3l}{2G}(\sigma+\frac{1}{\mu l}+\frac{\alpha(1-\alpha\Lambda_0 l^2)}{2\mu^2 l^2(1+\alpha\sigma)^2})
\end{equation}
Now by inserting left and right temperatures from Eq.(\ref{35}), and also the above central charges into Eq.(\ref{36}) we obtain
\begin{equation}\label{38}
S=\frac{\pi}{2G}(\sigma r_++\frac{r_-}{\mu l}+\frac{\alpha r_+(1-\alpha\Lambda_0 l^2)}{2\mu^2 l^2(1+\alpha\sigma)^2})
\end{equation}
which is exactly the same result Eq.(\ref{152}), we have obtained from general entropy formula for BTZ black hole solution of MMG.
\section{Conclusion}
In the present paper we have obtained the conserved current and corresponding charge for entropy of black hole solutions of MMG, which have been given by Eqs.(\ref{444'}), (\ref{11}) respectively. Equation (\ref{11}) give us the general conserved charge for calculations of
entropy of any black hole solution of MMG. Then according to the method of Tachikawa \cite{a} for calculation of entropy, we have used equation
(\ref{12}), and could obtain the entropy of BTZ black hole solution of MMG. Our result is given by Eq.(\ref{152}). As one can see, this expression contain 3 terms. The first term come from Einstein gravity in the presence of negative cosmological constant. The second term is due to the CS term, and last term is the contribution of new term of MMG, and is proportional to the parameter $\alpha$. The CS term in the Lagrangian of MMG is non-covariant and its contribution to the entropy is proportional to $r_-$, however, the cosmological Einstein gravity part and new term of MMG give us a contribution proportional to $r_+$.
After that in order to check that our result (\ref{152}) is correct, we have reproduced it by using the Cardy formula for entropy of boundary CFT. The result by this method is exactly coincide with (\ref{152}).
\section{Acknowledgments}
M. R. Setare  grateful to Y. Tachikawa,  G. Giribet,  M. H. Vahidinia, and G. Clement  for helpful discussions and correspondence.


\begin{thebibliography}{99}
\bibitem{a}Y. Tachikawa, Class. Quant. Grav. 24, 737, (2007).
\bibitem{17}E. Bergshoeff, O. Hohm, W. Merbis, A. J. Routh and P. K. Townsend,  Class. Quant. Grav. 31, 145008, (2014).
\bibitem{4}S. Deser, R. Jackiw and S. Templeton, Annals Phys.
140, 372 (1982) [Erratum-ibid. 185, 406.1988 APNYA,
281,409 (1988 APNYA,281,409-449.2000)].
\bibitem{18}A. S. Arvanitakis, A. J. Routh and P. K. Townsend, arXiv:1407.1264 [hep-th]; A. Baykal, arXiv:1408.5232
[gr-qc]; B. Tekin, arXiv:1409.5358 [hep-th]; G. Giribet, Y. Vásquez, arXiv:1411.6957 [hep-th].
\bibitem{f}M. R. Setare, arXiv:1412.2151 [hep-th].
\bibitem{6}E. A. Bergshoeff, O. Hohm and P. K. Townsend, Phys. Rev. Lett. 102, 201301, (2009).
\bibitem{e}M. Banados, C. Teitelboim and J. Zanelli, Phys. Rev. Lett. 69, 1849, (1992).
\bibitem{g}R. M. Wald, Phys. Rev. D48, 3427, (1993).
\bibitem{h}T. Jacobson, G. Kang and R. C. Myers, Phys. Rev. D49, 6587, (1994).
\bibitem{i}V. Iyer and R. M. Wald, Phys. Rev. D50,  846, (1994).
\bibitem{d}S. N. Solodukhin, Phys. Rev. D 74, 024015, (2006).
\bibitem{j}W. Kim, S. Kulkarni, S.H. Yi, Phys. Rev. D 88, 124004, (2013).
\bibitem{b}J. Lee and R. M. Wald, J. Math. Phys. 31,  725, (1990).
\bibitem{c}R. Wald, J. Math. Phys. 31,  2378, (1990).
\bibitem{k}A. Bouchareb, G. Clement, Class. Quant. Grav.24, 5581, (2007).
\bibitem{19}W. Li, W. Song, A. Strominger, JHEP 0804, 082, (2008).
\bibitem{20} A. Strominger, JHEP 9802, 009, (1998).
\bibitem{21}H. Saida, J. Soda,  Phys. Lett. B471, 358, (2000).
\end{thebibliography}
\end{document}